\documentclass[12pt,preprint]{aastex}

\shorttitle{}
\shortauthors{Young \& Arnett}

\newcommand{\sol}{$M_\odot$}
\def \nuc#1#2{\relax\ifmmode{}^{#1}{\protect\text{#2}}\else${}^{#1}$#2\fi}

\begin{document}

\title{Observational Tests and Predictive Stellar Evolution II: Non-standard Models}

\author{Patrick A. Young and David Arnett}
\affil{Steward Observatory, University of Arizona, 
933 N. Cherry Avenue, Tucson AZ 85721}
\email{payoung@as.arizona.edu, darnett@as.arizona.edu}

\begin{abstract}
 
We examine contributions of second order physical processes to the
results of stellar evolution calculations which are amenable to direct
observational testing. In the first paper in the series \citep{ymal}
we established baseline results using only physics which are common to
modern stellar evolution codes. In the current paper we establish how
much of the discrepancy between observations and baseline models is
due to particular elements of new physics in the areas of mixing,
diffusion, equations of state, and opacities. We then consider the
impact of the observational uncertainties on the maximum predictive
accuracy achievable by a stellar evolution code. The sun is an optimal
case because of the precise and abundant observations and the relative
simplicity of the underlying stellar physics. The Standard Model is
capable of matching the structure of the sun as determined by
helioseismology and gross surface observables to better than a
percent. Given an initial mass and surface composition within the
observational errors, and no current observables as additional
constraints for which the models can be optimized, it is not possible
to {\it predict} the sun's current state to better than $\sim 7\%$.
Convectively induced mixing in radiative regions, terrestrially
calibrated by multidimensional numerical hydrodynamic simulations,
dramatically improves the predictions for radii, luminosity, and
apsidal motions of eclipsing binaries while simultaneously maintaining
consistency with observed light element depletion and turnoff ages in
young clusters \citep{ykra}. Systematic errors in core size for models
of massive binaries disappear with more complete mixing physics, and
acceptable fits are achieved for all of the binaries without
calibration of free parameters. The lack of accurate abundance
determinations for binaries is now the main obstacle to improving
stellar models using this type of test.

\end{abstract}

\keywords{stars: evolution - stars: fundamental parameters -
  hydrodynamics - convection - lithium}

\section{INTRODUCTION}

Stellar evolution has become a successful tool for elucidating the
processes at work in individual stars. New instrumentation and a
wealth of new data has resulted in increased emphasis in astronomy on
the evolution of galaxies over cosmic history. Obviously the stellar
content of a galaxy plays a central role in its evolution. In order to
understand this process, we require theoretical stellar evolution to
be predictive, in the sense of being able to accurately describe the
contribution of luminosity, kinetic energy, and nucleosynthetic
products from a star of a specific initial mass and composition at any
and all points in its life. This process must be able to be carried
out self consistently for stars from the hydrogen burning limit to the
highest possible masses, so that stellar outcomes can be reliably
linked to an initial mass function (IMF). This is not possible with
schemes which are adjusted to match astronomical observations. Without
an underlying physical theory, the calibration cannot be reliably
extrapolated to regimes without extensive and independent
observational data. Also, multiple physical effects can contribute in
opposite or orthogonal senses to the star's structure. As an example,
determinations of metallicity of binaries are often made by fitting
model tracks of varying composition to observed points and finding the
best fit. However, more mixing produces more luminous models for a
given mass at a given effective temperature. Lower metallicity shifts
tracks bluewards and slightly higher in luminosity. Either change
could force a model to pass through the desired observed point. So a
model with incomplete mixing physics and solar composition could be as
good a formal fit to specific observations as a model with more
complete physics and supersolar metallicity.

The two primary areas which strongly affect the evolution and have
uncertain physics are mixing and mass loss. The evolution is also
sensitive to the opacity of the stellar material, but the opacities
used in evolution codes are largely uniform, utilizing the OPAL values
of \citet{ir96} for high temperatures and \citet{af94} for low
temperatures. The observational errors in determining stellar
compositions are the major limitation on further testing contribution
of opacities to stellar models. Even the metallicity of the sun varies
from z=0.019 \citep{gn93} to z=0.015 \citep{lo03}, depending upon the
precise physical assumptions and dimensionality of the codes being
used to fit the measured line profiles. Most other test cases, such as
most double-lined eclipsing binaries, {\em have no published
metallicity determinations using high resolution
spectroscopy}. Equations of state (EOS's) are not uniform across
stellar evolution codes. While the effects of the EOS are perhaps more
subtle, they can still be important, particularly for low mass stars
and pre- and post-main sequence evolution.

In \citet{ymal}, we presented baseline
results from stellar models calculated using only physics common to
current widely used stellar evolution codes. These models were tested
against a subset of double-lined eclipsing binaries
\citep{and91,rib00, lat96, las02, hw04}. \citet{ykra}
discussed hydrodynamic mixing within the radiative regions of stars
and presented several observational tests of the mechanism. This paper
presents a reanalysis of the eclipsing binary sample and solar models,
with more realistic mixing physics as well as additional minor
improvements to the code. Section 2 summarizes the additional physics
and improvements to the code. Solar models are examined in Section
3. The eclipsing binary sample is presented in Section 4. Section 5
contains discussion and conclusions. The implications for post-main
sequence evolution will be presented in a subsequent paper.

\section{THE TYCHO CODE}

The TYCHO code is a 1D stellar evolution and hydrodynamics code
written in structured FORTRAN77 with online graphics using PGPLOT. The
code is as described in \citet{ymal}. We have made substantial
additions and improvements. The code is now functional for stars from
the hydrogen burning limit to arbitrarily high masses, and for
metallicities of z = 0 to the limit of the OPAL opacity tables.

The opacities used are from \citet{ir96} at high temperatures and
\citet{af94} for low temperatures. The OPAL tables have been extended
to low entropies, and are formally adequate for calculating stellar
interiors down to the hydrogen burning limit. In reality, a number of
contributions, particularly from molecular species, are not
included. Stellar models computed with these tables are reliable to
perhaps 0.5 \sol.

TYCHO uses an adaptable set of reaction networks, which are
constructed automatically from rate tables given a list of desired
nuclei. In these calculations a 176 element network complete through
the iron peak was used at $T > 10^7$\ K, and a 15 element network for
light element depletion at lower temperatures. Rates for the full
network are from \citet{rt00}. \citet{cf88} rates are used in the
light element network.

Mass loss capabilities of the code have been extended. At $T_{eff} >
7.5\times 10^3$K, the theoretical approach of \citet{kud89} is
used. At lower $T_{eff}$\ routines based upon the empirical
prescription of \citet{rei} or \citet{blo95} are available. Low
temperature mass loss was not important in any of the cases studied
here, and the Reimers and Bl\"ocker algorithms converge in the limit
of low luminosity. A treatment of radiatively driven mass loss in
Wolf-Rayet stars based upon the work of \citet{ln03} is also included
in the code. It does not come into play for these models and will be
discussed in a separate paper.

The equation of state has been expanded from the modified \citet{ts00}
EOS in \citet{ymal} to include a more generalized treatment of the
coulomb properties of the plasma. The formation and dissociation of
molecular hydrogen and its effect upon the equation of state are also
included in a Helmholtz free energy formulation. The OPAL project has
extended its EOS determinations to lower entropies. The improved TYCHO
equation of state agrees with the OPAL EOS to better than 1\% for most
conditions. There remains a difficult region ($10^{-1} \leq \rho \leq
10^1 {\rm g\ cm^3}$ and $T \leq 10^{5.5}$ K) in which the plasma is a
strongly interacting coulomb system, and in which the difference
exceed 4\%. This region is relevant for low mass stars ($M < 0.5
M_\odot$).

TYCHO uses a modified version of Ledoux convection which avoids the
problem of instantaneous mixing in convective regions with nuclear
burning during short timesteps. Simple Eddington-Sweet rotational
mixing \citep{tas} is implemented in the code. Rotation was not
included in the binary models as these systems are all relatively slow
rotators for their mass range. Based on test models, rotation seems to
be a perturbation of the order of the observed errors or less on the
observed quantities we are examining.

All models were run with an improved version of the inertial wave
driven mixing described in \citet{ykra}. This treatment is primarily
derived from \citet{press81} and references therein. Press begins from
two separate approximations to describe the propagation of internal
waves in a stellar interior. The global behavior of the waves is
described by the anelastic approximation \citep{doux74}. Boussinesq,
plane equations with thermal diffusion describe the damping behavior
of linear internal waves \citep{lh78}. \citet*{press81} combines
these equations to define a set of globally anelastic but locally
Boussinesq equations. The principal equation of motion is

\begin{equation}
\frac{\partial^2{\bf \Psi}}{\partial r^2} + \left(\frac{N^2}{\omega^2} - 1\right)k_H^2{\bf \Psi} + \frac{i\sigma}{\omega}\left(\frac{\partial^2}{\partial r^2}-k_H^2\right)^2{\bf \Psi} = 0
\end{equation}

where

\begin{equation}
{\bf \Psi} \equiv \rho_0^{\case{1}{2}}k_H^{-2}u_V,
\end{equation}

$N$ is the composition dependent Brunt-V\"{a}is\"{a}l\"{a} frequency,
\begin{equation}
N^2 = -g\frac{1}{\rho_0}\frac{d\rho_0}{dz} - \frac{1}{\Gamma_1P}\frac{dP}{dr}, 
\end{equation}

$k_H$ is the local horizontal wavenumber,
\begin{equation}
k_H \equiv \left[\frac{l(l+1)}{r^2}\right]^{\case{1}{2}},
\end{equation}

$\sigma$ is the thermal diffusivity,
\begin{equation}
\sigma \equiv \frac{4(\Gamma1 - 1)}{\Gamma1}\left(\frac{\case{1}{3}aT_0^4}{P_0}\right)\left(\frac{c}{\kappa_0\rho_0}\right),
\end{equation}

and $\omega$ is the characteristic frequency of the driving, in this
case convective modes of low frequency. The vertical fluid velocity
$u_V$ (for moderate damping) is determined from a WKB solution of
equation 1, and the vertical wavenumber $k_V$ is defined as

\begin{equation}
k_V \equiv k_H\left[\frac{N^2}{\omega^2} - 1\right]^{\case{1}{2}}.
\end{equation}

The quality factor $Q$ of the propagating waves is

\begin{equation}
Q = 2\frac{N}{\sigma k_H^2}\left(\frac{\omega}{N}\right)^3\left(\frac{N^2-\omega^2}{N^2}\right).
\end{equation}

Waves generated by the convective zone are in most circumstances
non-linear within one linear damping distance of the convective
zone. Mixing in this region is efficient, but it is a small fraction
of the star. We are also interested in the impact of the waves outside
this zone. Damping of the waves in the radiative region generates
vorticity as described in \citet{ykra}. \citet*{press81} derives a mass
diffusivity coefficient $\sigma_M$, which we adopt, as,

\begin{equation}
\sigma_M \sim \frac{\epsilon^4\sigma^2k_V^2}{\omega} \sim \sigma\frac{\epsilon^4}{Q} 
 \end{equation}

where $\epsilon$ is the dimensionless nonlinearity parameter for internal waves

\begin{equation}
\epsilon \equiv \frac{k_Hu_H}{\omega}.
\end{equation}

The efficiency of the wave propagation, and therefore the mixing, is
inherently sensitive to composition gradients through $N$ and,
indirectly, through the opacity in $\sigma$. The mixing occurs roughly
on a thermal timescale. It is slow relative to convective turnover,
but it can be significant on evolutionary timescales. In practice,
this drives mixing in what are typically referred to as
``semiconvective'' regions, where $(\frac{N^2}{\omega^2}-1) > 0$. When
this quantity becomes imaginary, the wave undergoes complete internal
reflection, and there is an exponential falloff of the wave flux
beyond this boundary. In terms of composition gradients, there is a
region outside the convective zone which is well mixed, with a further
sharply decreasing degree of mixing beyond. This treatment removes the
one variable parameter from \citet{ykra}, the characteristic length
for dissipation, by treating the wave physics more completely. Our
results are little changed from the earlier formulation. This is
unsurprising, as we constrained the earlier treatment with numerical
simulations in which the wave behavior was readily apparent.

Gravitational settling and differential diffusion of
nuclear species according to \citet{tbl94} is also included. The
\citet{tbl94} treatment of diffusion is generalizable to an arbitrary
number of nuclear species, though that work examines only H, He, O,
and Fe. We calculate diffusion coefficients separately for the species
important to the OPAL ``type 2'' opacity tables (H, He, C, and O) and on
average for iron peak and Ne like elements. \citet{mic04} examine the
effect of settling on the approximately solar age and metallicity
clusters M67 and NGC 188, using 19 elements. In light of these results
our intermediate simplification appears adequate for stellar structure
calculations.

Numerous minor improvements have been made which improve convergence
and stability of the code, and allow it to perform adequately at the
small timesteps typical of neutrino-cooling dominated burning stages
as well as the slow hydrogen burning stages. The code is publicly
available and open source. The current version (TYCHO-7.0) is being
made available, along with an extensive suite of analysis tools, at
http://pegasus.as.arizona.edu/$\sim$darnett.

\section{SOLAR MODELS}

As the best observed star in the sky, the Sun is an obligatory test
case for any comprehensive stellar evolution code. The
helioseismological measurements of sound speed and depth of the
convective zone give us an insight into the interior structure not
available for any other star. In this section we test solar models
from TYCHO, but with a somewhat novel aim. We hope that TYCHO will
function as a predictive tool for building stellar
populations. Therefore, instead of finding a best fit to objects as
they are observed now, with variable initial conditions, the code must
be able to predict a unique (and accurate) path through stellar
parameter space over time for a particular initial mass and
composition. Conversely, we would also like to connect any given
observed star to a unique initial condition. We would wish
to do this for the complete range of stellar masses. As such we are
more interested in the comparison of our models with the sun, assuming
only an initial solar mass, composition, and our best treatment of
the physics, than in how precisely we can fit the sun by optimizing
our models. A 1 $M_\odot$ star on the main sequence is probably the
easiest type of star to model, being relatively insensitive to the
effects of mixing and mass loss. Solar models give us an estimate of
the minimum uncertainty in our predictions of stellar parameters.

We examine four models, s0, s1, s2, and s3 which differ in the
completeness of mixing physics included. Model s0 includes
gravitational settling and diffusion \citep{tbl94} and inertial
wave-driven mixing \citep{ykra}. Model s1 includes only wave-driven
mixing and s2 only gravitational settling and diffusion. Model s3 uses
only Ledoux convection and ignores other mixing
physics. Eddington-Sweet mixing is disabled, as it is a poor
description of the true angular momentum distribution in the sun. We
also calculate one model (l0) with physics identical to s0, but with
\citet{lo03} values for solar abundances. 

There is one glaring limitation, a free parameter which must of
necessity remain in this 1-D code. We choose a mixing length parameter
of $\alpha=2.1$, where $\alpha$ is the ratio of the mixing length to
the pressure scale height. This is in the same range as values deduced
from solar standard models ($\alpha$ = 2.05) \citep{bpb00} and
multidimensional simulations of the solar convective zone with
hydrodynamics and radiative transfer ($\alpha$ =
2.13)\citep{rob04}. Smaller values of the mixing length parameter
result in larger radii for the 1-D models. The mixing length parameter
is a shorthand for more complex physics, and is not guaranteed to be
constant over stellar mass or evolutionary state, so these constraints
apply only to sun-like stars.

Table~\ref{tbl-1} gives values for gross observables ($R, T_{eff},
L$), the rms difference in predicted and observed sound speed, depth of
convection zone, photospheric He and Li values, and central
temperature for each model. (Note that the observed Li abundance in
the sun is much lower than the meteoritic value. The meteoritic value
is assumed to more accurately represent the abundance in the
proto-solar nebula, and as such is used as the starting value for the
models.) We have not performed an inversion of the helioseismological
data through our model to obtain expected sound speeds for our
models. The values to which we compared are those calculated by
\citet{bpb01}. We have roughly twice as many Lagrangian mass points as
the online tables for the standard models, so we perform an
interpolation in our sound speeds to match the standard model
tables. Performing this direct comparison we find rms errors in
sound speed of $\sim 0.5\%$\ for our best models. Most of this
discrepancy can be attributed to our error in the solar
radius. (\citet{bpb01} find a 0.15\% rms error for a model with a
0.04\% difference in radius from their standard value.)

Table~\ref{tbl-2} gives neutrino fluxes for the models and a selection
of models from \citet{bp04}. The \citet{bp04} model BP04 is the
current standard solar model. BP04+ includes improvements in nuclear
physics, solar equation of state, nuclear physics and solar
composition. The entry ``Comp'' shows the effects of variant
composition alone, and ${\rm N^{14}}$ shows the effects of a change in
the ${\rm N^{14}}(p,\gamma){\rm O^{15}}$ rate alone.

\begin{deluxetable}{crrrrrrrr}
\tabletypesize{\scriptsize}
\tablecaption{Characteristics of Solar Models with Varied Mixing Physics \label{tbl-1}}
\tablewidth{0pt}
\tablehead{
\colhead{Model} & \colhead{$R/R_\odot$}   & \colhead{$\log T_e$} & 
 \colhead{$\log L/L_\odot$} &
 \colhead{rms $\delta c_s$ (\%)}  &\colhead{$R_{conv}/R$}     & \colhead{$X_{He}$} & 
 \colhead{$log\ \epsilon (^7Li)$} & \colhead{$T_c$} } 

\startdata
s0 & 0.993 & 3.765  & $3.80\times 10^{-3}$ & 0.56 & 0.714 & 0.242 & 1.14 & $1.560\times 10^7$   \\
s1 & 0.972 & 3.770  & $7.65\times 10^{-3}$ & 0.90 & 0.729 & 0.279 & 1.78 & $1.564\times 10^7$    \\
s2 & 0.985 & 3.767  & $5.93\times 10^{-3}$ & 0.50 & 0.718 & 0.240 & 3.05 & $1.562\times 10^7$    \\
s3 & 0.971 & 3.770  & $6.85\times 10^{-3}$ & 0.97 & 0.729 & 0.279 & 3.11 & $1.563\times 10^7$    \\
l0 & 0.944 & 3.777  & $9.21\times 10^{-3}$ & 2.24 & 0.672 & 0.252 & 4.89 & $1.558\times 10^7$    \\
standard model\tablenotemark{a} & 1.0 & 3.762 & 1.0 & 0.10 & 0.714 & 0.244 & 1.1 & $1.569\times 10^7$   \\
\enddata 
\tablenotetext{a}{Solar values from standard solar model of \citet{bpb01} except $log\ \epsilon (^7Li)$ from \citet{bs03}}
\end{deluxetable}

\begin{deluxetable}{crrrrrrrr}
\tabletypesize{\scriptsize}
\tablecaption{Neutrino Fluxes for Solar Models with Varied Mixing Physics \label{tbl-2}}
\tablewidth{0pt}
\tablehead{
\colhead{Model} & \colhead{$pp$}   & \colhead{$pep$} & 
 \colhead{$hep$} &
 \colhead{${\rm Be^7}$ }  &\colhead{${\rm B^8}$}     & \colhead{${\rm N^{13}}$} & 
 \colhead{${\rm O^{15}}$} & \colhead{${\rm F^{17}}$}\\
\colhead{} & \colhead{$10^{10} {\rm cm^2 s^{-1}}$} & \colhead{$10^{8} {\rm cm^2 s^{-1}}$} & \colhead{$10^{3} {\rm cm^2 s^{-1}}$} & \colhead{$10^{9} {\rm cm^2 s^{-1}}$} & \colhead{$10^{6} {\rm cm^2 s^{-1}}$} & \colhead{$10^{8} {\rm cm^2 s^{-1}}$} & \colhead{$10^{8} {\rm cm^2 s^{-1}}$} & \colhead{$10^{6} {\rm cm^2 s^{-1}}$} }

\startdata
s0 & 5.95 & 1.42  & 7.91 & 4.83 & 5.51 & 4.08 & 3.49 & 4.59   \\
s1 & 5.97 & 1.42  & 7.84 & 4.86 & 5.59 & 4.12 & 3.53 & 4.65    \\
s2 & 5.96 & 1.42  & 7.84 & 4.81 & 5.44 & 4.04 & 3.45 & 4.54   \\
s3 & 5.97 & 1.42  & 7.85 & 4.84 & 5.51 & 4.09 & 3.49 & 4.58   \\
 & & & & & & & & \\
BP04\tablenotemark{a} & 5.94 & 1.40 & 7.88 & 4.86 & 5.79 & 5.71 & 5.03 & 5.91 \\
BP04+ & 5.99 & 1.42 & 8.04 & 4.65 & 5.02 & 4.06 & 3.54 & 3.97 \\ 
Comp & 6.00 & 1.42 & 9.44 & 4.56 & 4.62 & 3.88 & 3.36 & 3.77 \\
${\rm N^{14}}$ & 5.98 & 1.42 & 7.93 & 4.86 & 5.74 & 3.23 & 2.54 & 5.85 \\

\enddata 
\tablenotetext{a}{Neutrino fluxes from standard model and models with various improvements in physics from \citet{bp04}}
\end{deluxetable}

The values in Table~\ref{tbl-1} illustrate some of the subtleties
involved in distinguishing between models. If we accept a constraint
on the mixing length from simulations or helioseismology, all variants
of the model predict gross observables to within 3\%. The models with
more complete mixing physics show a slightly better agreement, but the
variation is less than the uncertainty in the exact nature of the
convection. The minimum uncertainty in our predictions must be take to
be larger than 3\%, because the error is dominated by a fictitious
parameter. Varying the mixing length by 0.1 results is roughly a 1\%
change in the radius. Simulations of red giant atmospheres
\citep{asi00} and observations of pre-main sequence (pre-MS) binaries
\citep{hw04, sta04} indicate that stars with low surface gravities and
larger convective cell sizes and/or Mach numbers and turbulent
pressures have different convective physics than main sequence stars
of the same luminosity. In a 1D description of convection, this
manifests as a change in the mixing length to values of roughly
1.5. Without constraints on the nature of convection the minimum
predictive uncertainty is roughly 7\% for a 1$M_\odot$ star of solar
age. Varying the abundance from \citet{gn93} to \citet{lo03}
introduces a further uncertainty of $\sim$5\% in surface observables,
and a much larger variance for helioseismological and abundance
tests. Model l0 is a poor fit primarily because the opacity at the
base of the convective zone is too low. \citet{pr81} find that
hydraulic enhancement of the photon diffusivity by wave motion acts
in a radiatively stable stratified region to increase the effective
opacity, so the Lodders abundances may not be irreconcilable with
solar observations. This may also bear upon the perennial problem of
sound speed discrepancies between models and helioseismology
immediately below the solar convection zone. The implications of this
process will be discussed in a subsequent paper.

Using helioseismology and detailed chemical abundances, we can begin
to discriminate between models. Unsurprisingly, model s3, with mixing
limited to Ledoux convection, is ruled out immediately because its He
and Li abundances are much higher than observed, essentially unchanged
from their primordial values. Model s1, with radiative region mixing
but no heavy element diffusion, is also eliminated by the size of the
convective zone and surface helium abundance. Clearly gravitational
settling and diffusion are necessary to fit the observed sun. The only
observable difference between the remaining diffusion only and more
realistic mixing models lies in the predicted photospheric Li
abundance. This is exactly what is to be expected, since
helioseismology tells us that, while mixing must be present in the
radiative regions to account for observed abundances, it cannot have a
large {\it structural} effect.

\citet{mic04} confirm this result for the solar age and metallicity
clusters M67 and NGC 188. They find that models with little or no
``overshooting'' are consistent with the observed color-magnitude
diagrams of the clusters. Our theory of mixing naturally predicts
little structural effect on solar type stars and only a small increase
in core size for stars which start with small convective cores. We
do see a significant effect on the sun during the pre-MS, when the
transient convective core established during partial CN burning is at
its largest.

With no mixing (save settling) outside the convection zone, model s2
greatly under-predicts the depletion of Li at the solar
photosphere. The model with complete mixing gives an abundance much
closer to the observed value of $log\ \epsilon (^7Li) = 1.1 \pm 0.1$
\citep{bs03}, though this, too is sensitive to the mixing length at
the factor of 2 to 4 level. The role of rotation coupled to
oscillations in driving mixing has been discussed extensively by many
investigators (\citep[i.e.][]{cdp95,pin02}). The work of
\citet{ct99,tkz02} suggests that mixing is damped in rotating stars on
the red side of the Li dip, corresponding to early G stars. If the
pre-MS sun was a slow rotator, it may be a limiting case where angular
momentum transport produces a minimal modification in the stellar
g-mode oscillation spectrum and mixing is at a maximum.  This provides
a possible explanation for the strong depletion of Li in the sun
relative to field G stars, and for the wide observed range of
depletions, but the problem is beyond the scope of this paper.

The values for neutrino fluxes in Table~\ref{tbl-2} all fall within
the range of variation found by \citet{bp04} for variant models with
improved physics. A selection of the \citet{bp04} models illustrating
the range of variation between the models are given in
Table~\ref{tbl-2}. The neutrinos do not provide a constraint on the
models at this level, but do confirm that none of the physics included
in the models is in conflict with the observations.

The solar models highlight some of the problems in assessing the
predictive power of stellar evolution codes. The models presented here
would be indistinguishable for a G2 star outside the solar system. The
errors in the gross observables could be compensated for by a change
in the mixing length without including the necessary physics of He and
heavy element diffusion and non-convective mixing. Helioseismology
would not be available to falsify such a model based upon convective
zone depth or sound speeds. Stars with abundance determinations from
high resolution spectroscopy are relatively few and far between. One
may argue whether it is then important to have complete
physics. Figure 1 illustrates the potential traps that lie in
validating code with a narrow selection of observations. At the solar
age models s0 and s2 (both with diffusion) are nearly identical, as
are s1 and s3 (both without). The effects of diffusion clearly
dominate in determining the sun's position in the HR diagram. On the
pre-MS during partial CN burning in the transient convective core, the
case is very different. Models s0 and {\it s1} (inertial wave-driven
mixing) are nearly identical, as are s2 and s3 (convective mixing
only). The shape of the pre-MS is determined primarily by the change
in convective core size resulting from including more complete mixing
physics, Diffusion has had insufficient time to make much
difference. In short, the evolutionary history is not unique. A model
which fits the present day sun perfectly may be substantially
inaccurate for other evolutionary stages (or equivalently mass ranges
or compositions) where different physics come into play. Calibrations
based upon any one type of data set should not be extended into other
regimes unless based upon a valid physical theory.

\begin{figure}
\figurenum{1}
\plotone{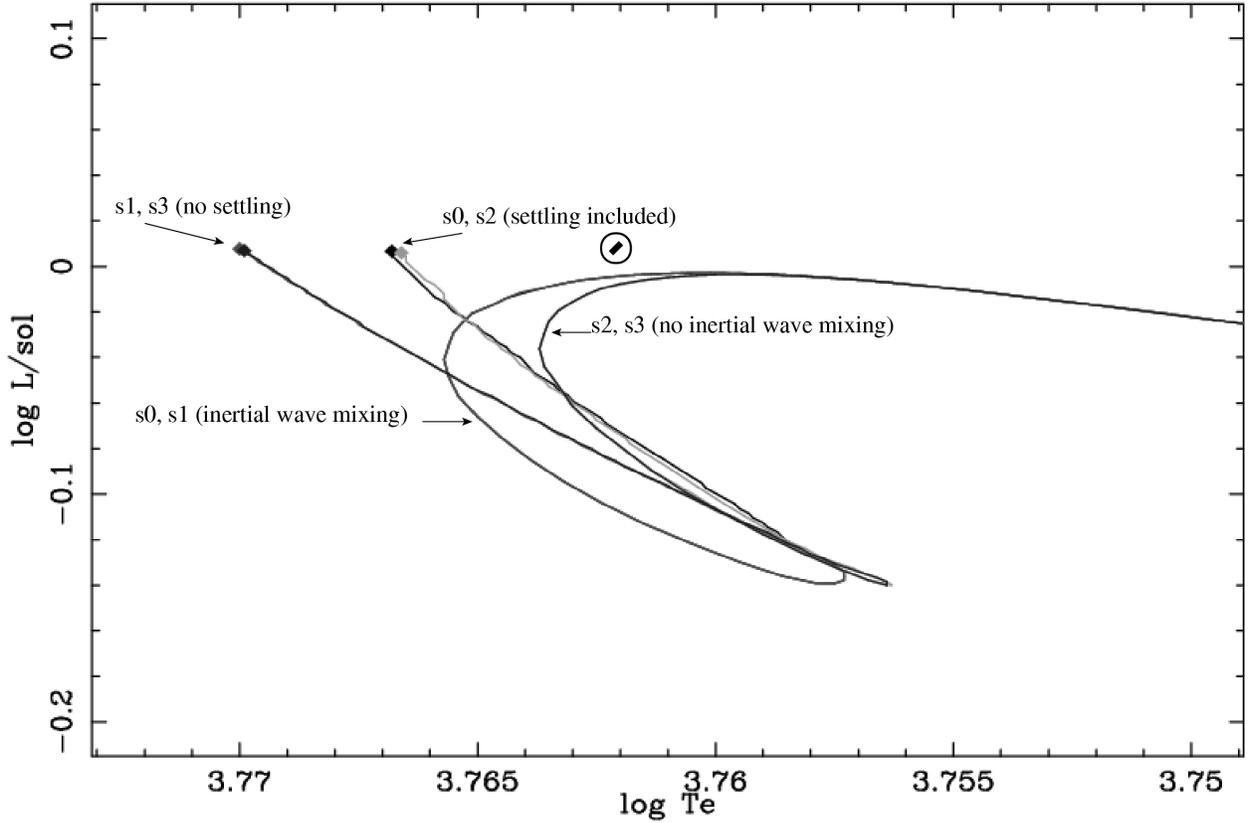}
\caption{Evolutionary tracks for 1 $M_\odot$ stars with four
variations on mixing physics. Model s0 includes gravitational
settling/heavy element diffusion and inertial wave-driven mixing in
radiative regions. Model s1 contains only wave mixing. Model s2
includes only settling/diffusion. Model s3 has no mixing outside of
convective regions. Models with hydrodynamic mixing in radiative
regions (s0 and s1) are indistinguishable on the pre-MS, when the
shape of the track is controlled primarily by the size of the small
convective core that exists during partial CN burning. Diffusion has
insufficient time to work on the pre-MS and does not affect the
tracks. On the main sequence, after the convective core disappears,
the shape of the track is determined by the presence (s0 \& s2) or
absence (s1 \& s3) of diffusion and settling.}
\end{figure}
\placefigure{fig1}

\section{ECLIPSING BINARIES}

The double-lined eclipsing spectroscopic binaries in \citet{and91}
provide us with a sample of stars from $1 M_\odot < M < 23 M_\odot$\
with precisely determined masses and radii. A subset of these stars
also have measured apsidal motions, which provide some information on
the interior density profiles and core sizes of the stars. We use the
same sample as \citet{ymal} so that a direct comparison of the same
code with and without complete descriptions of mixing physics outside
of convective regions can be made. We use the most recent values
available for observed quantities (observed quantities and references
can be found in Table~\ref{tbl-3}.)

As in \citet{ymal}, we calculate a $\chi^2$-like quantity for each
binary pair, defined by
\begin{eqnarray}
\chi^2& = &((\log L(m_A,t)-\log L_{A})/\sigma_{L}(A))^2\nonumber \\
 &+& ((\log L(m_B,t)-\log L_{B})/\sigma_{L}(B))^2 \nonumber\\
&+&((\log R(m_A,t)-\log R_{A})/\sigma_{R}(A))^2 \nonumber\\
&+&((\log R(m_B,t)-\log R_{B})/\sigma_{R}(B))^2,
\end{eqnarray} 
where $A$ and $B$ denote the primary and the secondary star,
respectively. Here $L_A$ and $R_A$ are the observationally determined
luminosity and radius of the primary, with $\sigma_{L A}$ and
$\sigma_{R A}$ being the observational errors in $\log L_A$ and in
$\log R_A$.  We convert the observational data for the radii to
logarithmic form for consistency. Correspondingly, $L(m_A,t)$ and
$R(m_A,t)$ are the luminosity and radius of the model.  This $\chi^2$
was evaluated by computing two evolutionary sequences, one for a star
of mass $m_A$ and one for $m_B$. A $\chi^2$ was calculated at
consistent times through the entire sequence ($t_A=t_B$ to a fraction
of a time step, which was a relative error of a few percent at
worst). The smallest $\chi^2$ value determined which pair of models
was optimum for that binary.  Note that if the trajectories of both
$A$ and $B$ graze the error boxes at the same time, $\chi^2\approx
4$. We use radius instead of effective temperature in our fitting
algorithm because the more precise values for $R$ make the $\chi^2$
more discriminating. These error parameters along with the
corresponding uncertainties from the observations are presented in
Table~\ref{tbl-4}.

As before, we note that a $\chi^2$ statistic assumes that the
observational errors have a Gaussian distribution about the mean
\citep{pre92}. This is not necessarily true, as the systematic shifts
in measured quantities due to new analyses can be much larger than the
formal error bars \citep{rib00, skp92, slc94}. Also, the quoted
luminosity depends on the radius and effective temperature, and is
thus not entirely independent. These systematic errors are the true
limit to our power to discriminate between models, and emphasize the
need for independent observational tests and numerical simulations to
identify relevant physics.

{\em Probably the greatest observational limitation we face is the
lack of abundance determinations for these stars.} The only binary in
our original sample with a spectroscopic abundance determination is UX Men
($z$=0.019)\citep{and89}. A few other systems have some sort of
metallicity indication in the literature. \citet{rib00} derive a
metallicity of $z$=0.013 from fits to evolutionary tracks for the ever
troublesome $\zeta$ Phe. Synthetic BaSeL photometry of VV Pyx suggests
a metallicity of $z<$0.007, but the fits are not good
\citep{las99}. \citet{lat96} and \citet{las02} argue for a metal
content similar to the Hyades in DM Vir ($z$=0.23). MY Cyg A\&B and AI
Hya A are all peculiar metal line stars. We (somewhat arbitrarily)
also assign these systems a Hyades composition. AI Hya has a measured
$z=0.07$, but this is probably a surface enhancement and does not
reflect the global composition of the star \citep{rib00}. Other
systems either have no metallicity determinations or are sufficiently
near solar composition that models of solar composition fall within
the observational errors.

\subsection{Global Properties of the Errors}

The $\chi^2$ values for each binary pair with and without complete
mixing physics are plotted in Figure 2. An arrow shows the shift in
$\chi^2$ from baseline models to models with the full suite of
physics. Fifteen of the binaries have excellent fits ($\chi^2 <
4$). Only three systems are marginal ($\zeta$ Phe, $\chi^2 = 4.10$; AI
Hya, $\chi^2 = 4.12$; EK Cep, $\chi^2 = 5.97$). These systems will be
discussed individually later. In all cases where the previous fits
were marginal to poor, the $\chi^2$ improved. All massive binaries
($M_A, M_B > 4 M_\odot$) with good fits also improved. Results were
mixed for lower mass stars with good fits. In both of the latter
groups, both complete and incomplete models fall within the
observational errors, so there is little to distinguish between models
for individual binaries. The threshold for rejection of a model with
$\nu = 4$ degrees of freedom with an $\alpha$ = 5\% chance of rejecting a
true hypothesis is $\chi^2 = 9.488$, so even the marginal fits do
not give us a strong hold on remaining errors in our models. We must
examine systematic discrepancies or wait for tighter observational error bars.

We show $\chi^2$ values for our best estimate of the metallicity. It
is difficult to make a global statement about how metallicity affects
the results of the tests, because it varies depending upon the
characteristics of the binary. For the most massive binaries, a change
in composition from solar ($z=0.0189$) to $z=0.004$ (roughly that of
the Large Magellanic Cloud) results in a shift in the HR diagram
roughly comparable to the observational errors. A similar shift in
abundances for the lowest mass stars can shift the models well outside
the region of good fit, primarily because of the smaller size of the
observational errors rather than an intrinsic physical
sensitivity. The converse is not necessarily true, however. AI Hya is
the most extreme case. Using a solar composition model with complete
mixing physics results in a $\chi^2 =11.2$, because the model is too
luminous at the correct $T_{eff}$. The converse situation does not
hold. A solar composition model with only standard physics has $\chi^2
> 20$. A supersolar composition does not produce a significant
improvement because the main sequence never travels far enough to the
red for either case. Only increased mixing solves this problem. In
general, it is safe to say that a change in metallicity of a factor of
2-4 is sufficient to move a binary model out of good agreement with
observations. A similar change may or may not make poor model fit
well. In order to pin down the stellar physics we must either examine
systematic trends or find better abundance determinations.

\begin{figure}
\figurenum{2}
\plotone{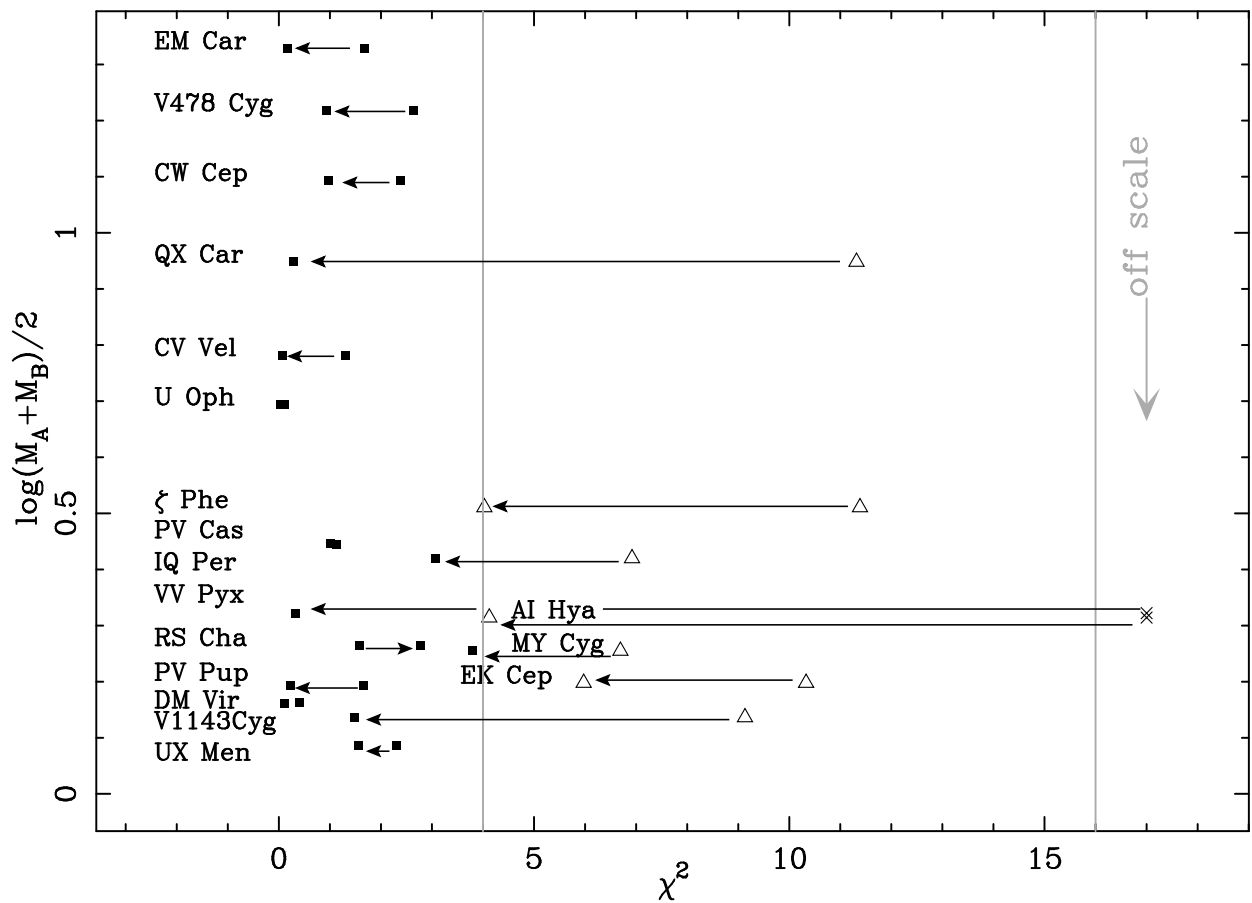}
\caption{$\chi^2$ values for optimum models of binaries, versus mean
mass of the binary. Arrows are drawn from \citet{ymal} values to
values for current models with more complete physics. The improvement
is dramatic. The vertical line at $\chi^2 = 4$ indicates a fit in
which the models are just within the observational errors for both
stars.}
\end{figure}
\placefigure{fig2}

Figure 3 shows goodness of fit vectors with the HR diagram for all
stars in the sample, with incomplete models on the left and complete
models on the right. The observed points with error bars are plotted
with an arrow indicating distance and direction to the best fit model
point. We can now begin to discriminate between models even for
formally excellent fits. The most striking feature of the figure is
the behavior of the massive stars ($M > 4 M_\odot$). The incomplete
models are systematically underluminous. This suggests three
possibilities: (1) the massive stars are all low metallicity; (2) the
observational luminosity and/or mass determinations are systematically
low; (3) the stars have larger convective cores than standard models
predict. Option (1) is unlikely for nearby massive stars with ages of
less than $10^8$ years, but cannot be absolutely ruled out without
spectroscopic abundance determinations. Option 2 is possible, but
again unlikely for a sample of 6 widely separated binaries. (Both EM
Car and CW Cep have had their masses revised {\it downward} by
\citet{skp92}.) Option 3 seems the most likely, and is consistent with
other evidence of mixing in stars beyond the standard model. Indeed,
the trend virtually disappears when realistic mixing is included. It
may be argued that the standard model is hydrodynamically
inconsistent, as indicated by the instantaneous deceleration required
at convective boundaries, and by detailed analysis of realistic high
resolution 3-D hydrodynamic simulations, problems which our mixing
algorithm addresses.

\begin{figure}
\figurenum{3}
\includegraphics[scale=1.0]{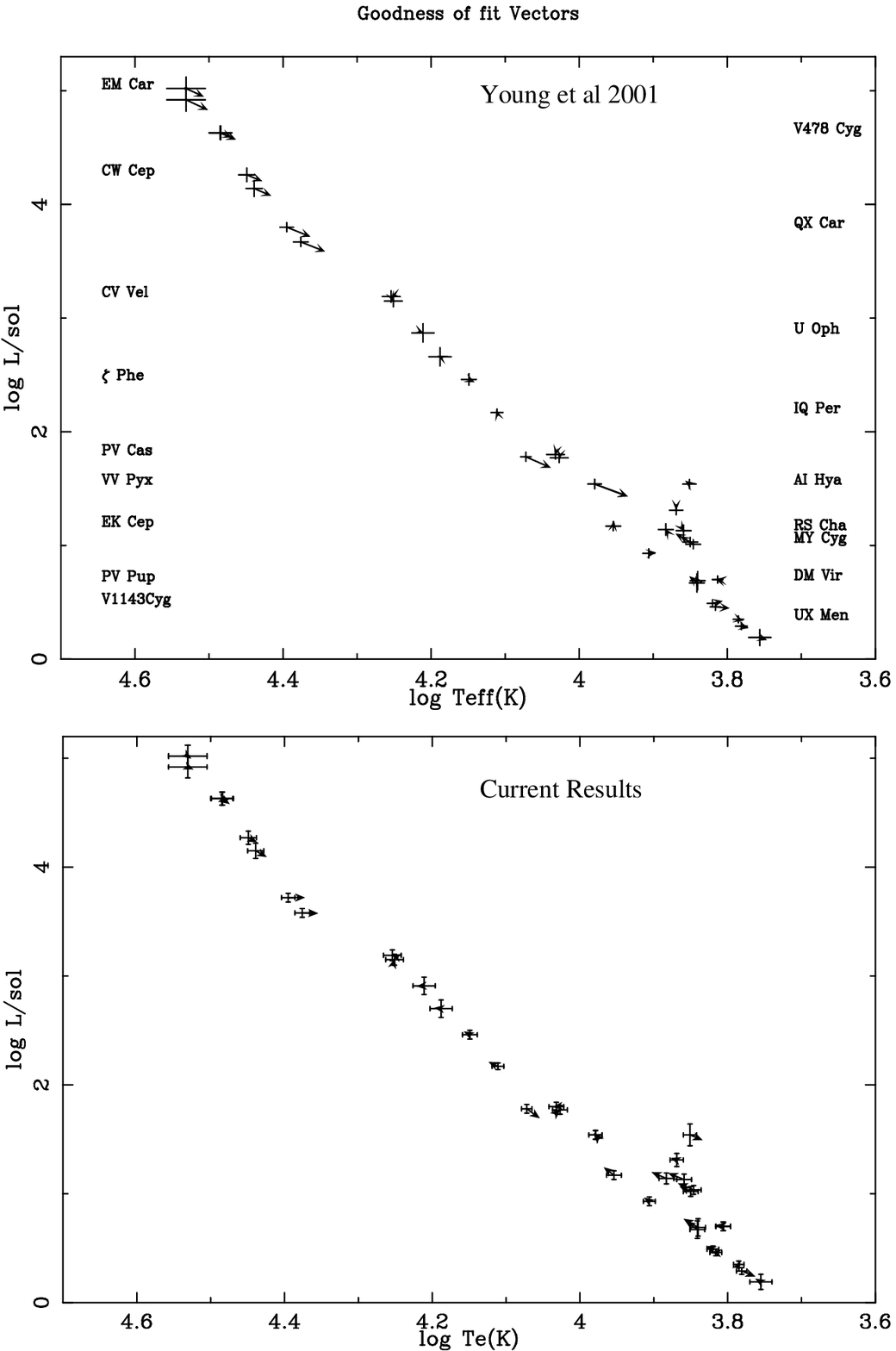}
\caption{Goodness of fit vectors for each member of binary sample,
with observational error bars. \citet{ymal} results are on the top;
present results on the bottom.}
\end{figure}
\placefigure{fig3}

The salubrious effect of realistic mixing is confirmed by the apsidal
motion tests. Models with incomplete mixing have systematically higher
predicted apsidal motions in the four most massive systems. This
indicates that the models are not sufficiently centrally
condensed. (See \citet{ymal} for a complete discussion of our previous
results and methodology.) The current models with realistic mixing
physics have larger convective cores and are therefore more centrally
condensed. Table~\ref{tbl-5} summarizes the apsidal motion results for
our binaries. Figure 4 shows the dimensionless rate of apsidal motion,
$ (P/U)_{CL} = (P/U)_{OBS} - (P/U)_{GR} $, which would be due to
classical apsidal motion, plotted versus log of half the total binary
mass (where $CL$ denotes classical and $GR$ general relativistic
parts, and $OBS$ the observed motion). $P$ is the orbital period and
$U$ the apsidal period.  The observational data (corrected for general
relativity) are shown as diamonds, with vertical error bars. All of
the massive star models now fall within the error bars for the
measured apsidal motion with the exception of QX Car, which differs by
roughly two and a half sigma (insofar as sigma is a meaningful
expression of these errors). QX Car does not differ from the measured
point by a larger absolute amount than the other binaries; it simply
has much tighter error bars. Either the observational uncertainties
are underestimated or, equally likely, the models are still missing
some physics. We have reanalyzed all of the binaries from EM Car to IQ
Per. The lower mass binaries all have quoted apsidal motions smaller
than the predicted general relativistic term. We find it more likely
that there are errors in measuring an apsidal motion with periods of
centuries than in weak-field general relativity.

\begin{figure}
\figurenum{4}
\plotone{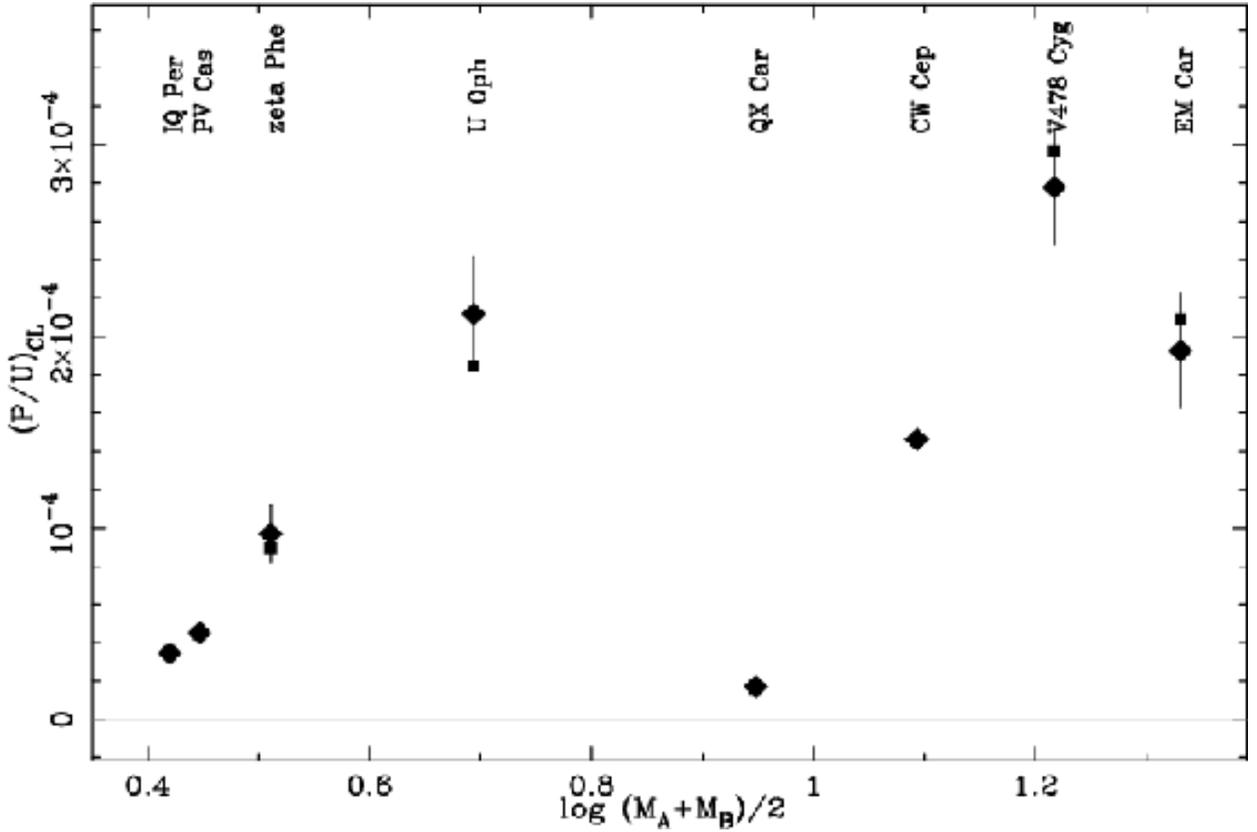}
\caption{Classical apsidal motion versus mean mass, for our binaries
with measured apsidal motion. $ (P/U)_{CL} = (P/U)_{OBS} - (P/U)_{GR}
$ is assumed. Diamonds are observed values, squares are model
fits. Note that the symbol for QX Car is larger than error bars, and
QX Car is not a good fit.}
\end{figure}
\placefigure{fig4}

\subsection{Individual Systems of Interest}

\subsubsection{$\zeta$ Phe}

This system is perennially troublesome to stellar modelers
\citep[i.e.][]{rib00}. It is difficult to fit both components with the
same metallicity. We adopt $z = 0.013$, following \citet{rib00} and
achieve a marginal fit ($\chi^2 = 4.026$). The secondary star is more
luminous than the models when a good fit is achieved for the
primary. If the observations are correctly interpreted, then the
secondary model requires either a lower metallicity or enhanced
mixing. Figure 5 shows the observed points and model tracks for
$\zeta$ Phe.

\begin{figure}
\figurenum{5}
\plotone{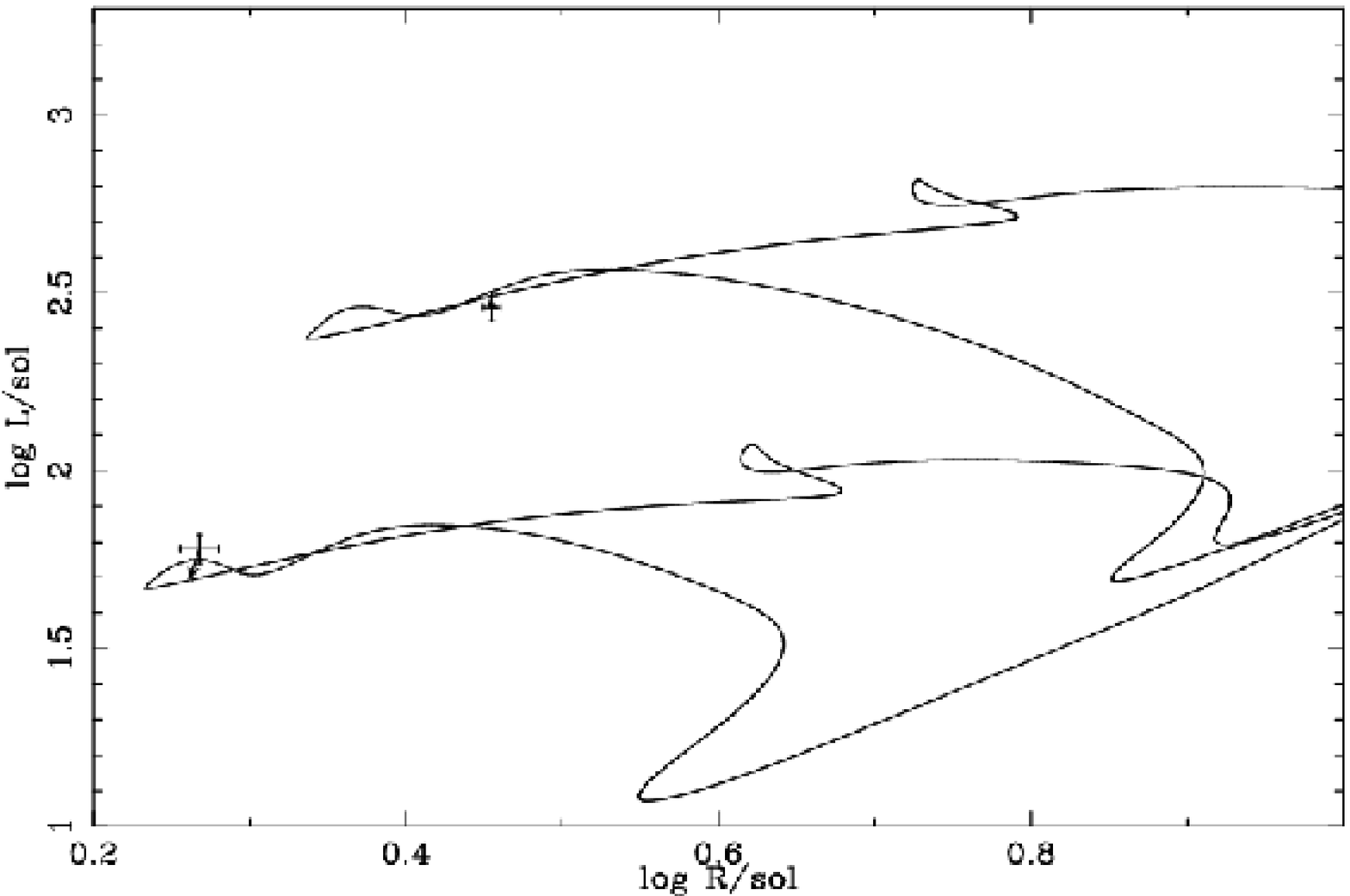}
\caption{Evolutionary tracks ($z = 0.013$) and observed points with
errors for $\zeta$ Phe. Note that the secondary star is more luminous
than the models when the primary is well fit.}
\end{figure}
\placefigure{fig5}

\subsubsection{AI Hya}
AI Hya is identified as a peculiar metal line star (spectral class
F2m). Its metallicity is measured as $z = 0.07$ \citep{rib00}, but
this is probably due to a surface enhancement. Still, the stars are
probably metal rich relative to solar. Without a precise determination
of the interior composition we choose to use a Hyades composition ($z
= 0.023$) as being in the reasonable range of nearby metal rich
compositions. In keeping with our effort to test the predictability of
our code, we do not try to optimize the fit by further varying the
composition. This system is particularly interesting in that increased
metallicity alone cannot reconcile tracks with only convective mixing
with the observations. The primary of the system lies farther redward
in the HR diagram than the terminal age main sequence
\citep{ymal}. The lifetime on the Hertzsprung gap for a 2 $M_\odot$
star is short. It is possible to catch a star in that stage, but
unlikely, as $\tau_{gap} / \tau_{MS} < 2\%$. More realistic mixing
gives a larger convective core, extending the track redward so that a
fit on the main sequence is easily achievable. Figure 6 shows the
observed points and model tracks for AI Hya.

\begin{figure}
\figurenum{6}
\plotone{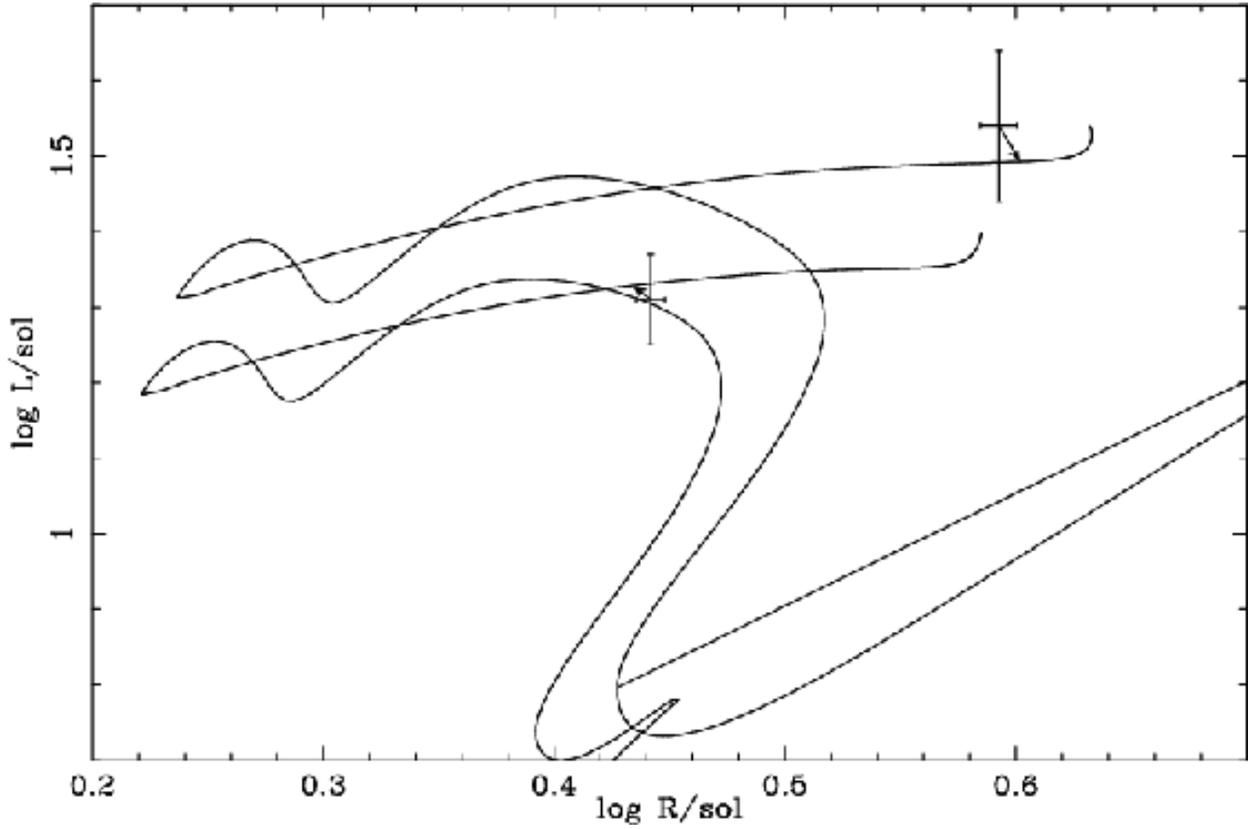}
\caption{Evolutionary tracks ($z = 0.023$) and observed points with
errors for AI Hya. Realistic mixing allows an acceptable fit to this
previously difficult pair.}
\end{figure}
\placefigure{fig6}

\subsubsection{EK Cep}
EK Cep and RS Cha are both pre-MS systems \citep{pop87, mlf00}. The
fit to RS Cha is formally a good one ($\chi^2 = 2.776$), and we do not
attempt to optimize within the observational errors. EK Cep, however,
achieves only a marginal fit ($\chi^2 = 5.973$), because the radius of
the secondary star is larger that that of the models during the first
rise of the CN burning bump. (Other early pre-MS models show similar
behavior, so we suspect this is systematic. A larger sample will be
discussed in a forthcoming paper.) This is a robust behavior, in the sense
that most things we could do to the models do not push them in the
right direction. The non-convective mixing physics does not have a
substantial effect, and increased metallicity would change the
luminosity too much to result in a good fit. We can only find good
agreement by reducing the mixing length parameter to $\alpha =
1.6$. This suggests a change in the nature of the convection, but
since $\alpha$ does not represent a physical entity, it does not tell
us what that change is. We may speculate that since the star is trying
to transport an amount of energy to a surface with a larger radius
than a main sequence star of similar luminosity, the convective Mach
numbers must be higher. This contributes a proportionally greater term
to the stress tensor than main sequence convection and manifests as a
radially directed pressure term, which would result in a larger radius
for hydrostatic equilibrium. Besides this term, there are plasma
effects, non-hydrogenic molecular contributions to the EOS, molecular
and grain opacities, and subtleties of atmosphere models which must be
taken into account which may contribute to the resolution of the
problem. In short, we can identify a deficiency in our physics, and
probably localize it to the physics of convection, but we do not have
good predictive accuracy in this evolutionary stage. We quote the
$\chi^2$ for our usual value of $\alpha = 2.1$, and not the improved
fit for $\alpha = 1.6$, since this is not a predictable change. We
plan 3-D simulations of convection in pre-MS and MS stars, which we
hope will characterize the difference in the convection in a physical
way. Figure 7 shows the observed points and model tracks for EK Cep.

\begin{figure}
\figurenum{7}
\plotone{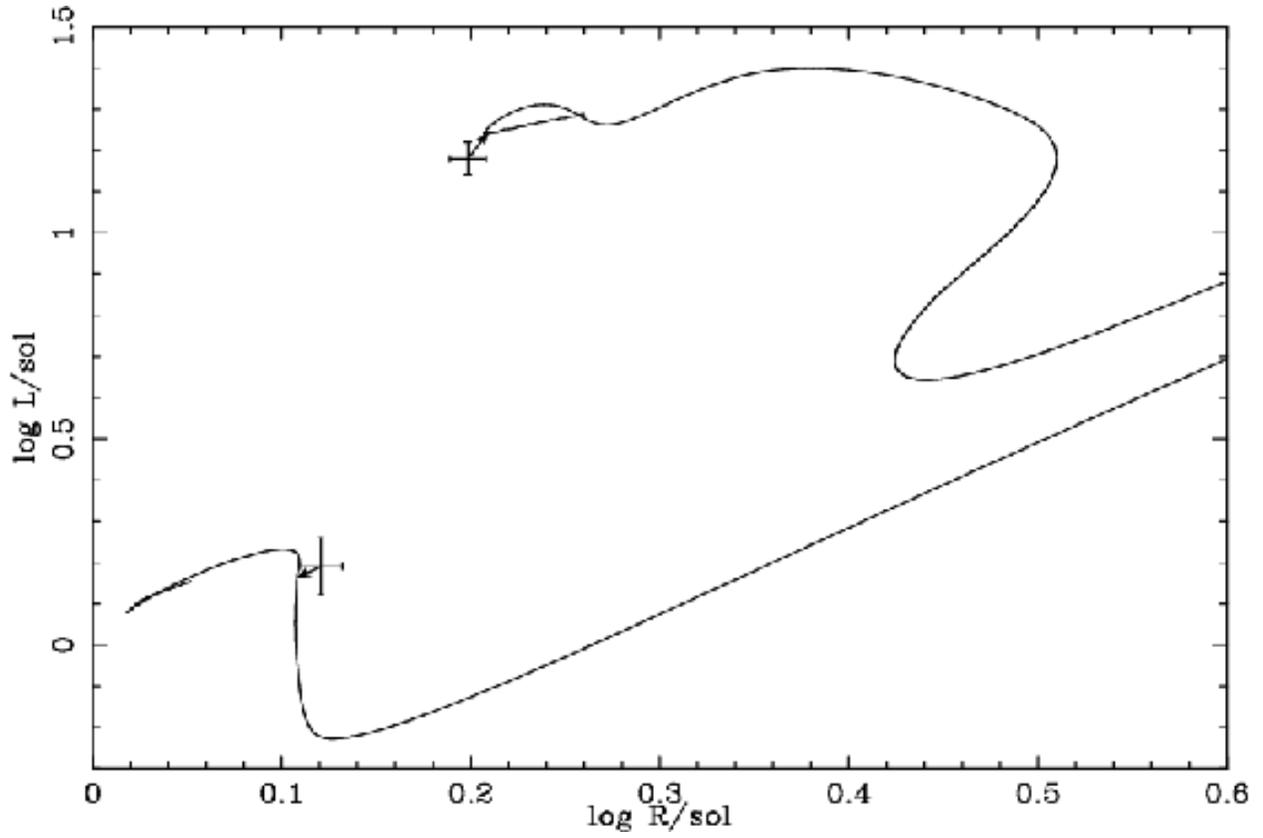}
\caption{Evolutionary tracks ($z = 0.019$) and observed points with
errors for EK Cep. The models for the secondary star were run with a
reduced mixing length parameter of $\alpha = 1.6$ in order to increase
the model radius to that of the observations. This indicates a change
in the nature of the convection.}
\end{figure}
\placefigure{fig7}

\subsubsection{TZ For}

TZ For was not in our original binary sample, but is a sufficiently
interesting system that we examine it briefly here. The secondary is a
subgiant in the Hertzsprung gap \citep{pol97} with a spectroscopically
determined metallicity of $z = 0.024 \pm 0.007$. \citet{las02} attempt
to fit the secondary with several stellar evolution codes, but are
unsuccessful without changing the mass of the model by 5$\sigma$ or
using a composition not in agreement with the observations.Figure 8
shows the observed points and model tracks for TZ For.

Changing the size of the convective core on the main sequence
necessarily changes the path the star takes across the Hertzsprung
gap. We find that with realistic mixing, our models match the hotter
component of TZ For reasonably well, and the cooler component
exceptionally well. The models for the subgiant are slightly
overluminous. We find a $\chi^2 = 0.77$ for the binary at $t =
5.3\times 10^8$ yr. Virtually all of this discrepancy comes from the
subgiant. 

\begin{figure}
\figurenum{8}
\plotone{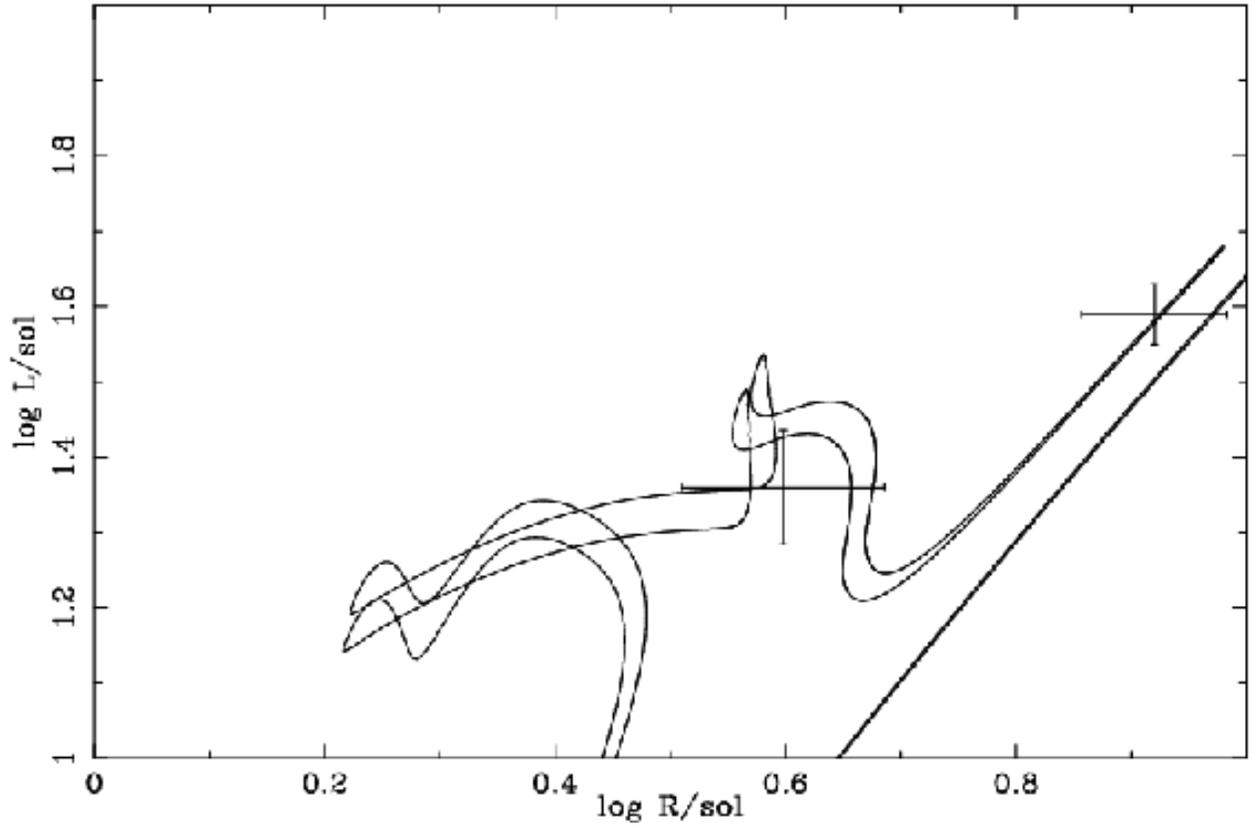}
\caption{Evolutionary tracks ($z = 0.024$) and observed points with
errors for TZ For. The models are slightly overluminous with respect
to the observed hot component, but much improved over standard
models.}
\end{figure}
\placefigure{fig8}

\section{CONCLUSIONS}

In this paper we test the predictive power of the TYCHO stellar
evolution code against a set of classical observational tests. With an
improved version of the realistic mixing physics presented in
\citet{ykra}, we find excellent agreement with solar models and the
sample of double-lined eclipsing binaries from \citet{ymal}. By
avoiding optimization of our models with composition changes or
parameterized extra mixing, we also identify several issues which are
important to future development of stellar modeling.

From the solar models we find that our predictive accuracy for the
sort of gross observables ($L,R,T_{eff}$) we measure for other stars
is limited to of order 5-10\% by (1) inadequacy in our description of
convection, manifested by an uncertainty in the fictitious mixing
length parameter, and (2) by uncertainties in abundances. If the
nature of the convection is fixed by numerical simulations of full 3-D
convection, the uncertainty is reduced to that arising from the
abundance determinations. The good agreement of the neutrino fluxes
with those of the standard model indicate that the influence of the
mixing length description is an ``atmospheric'' effect. The rest of our
(more or less parameter-free) physics provides a good description of
the interior of the sun. We wish to emphasize that the quoted
uncertainty is not a precise figure, but rather our best estimate,
accounting for various known errors and incomplete physical
descriptions. The sun is a well constrained system, and has
uncomplicated physics compared to most other stars. In this light our
measure of uncertainty should be taken as a best case for stellar
evolution, with the caveat that errors can become substantially worse
as we examine more complex, less well observed systems.

One of the most striking features of the solar models underlines a
fundamental problem of stellar evolution. The two models that match
observed solar quantities best have virtually identical tracks on the
main sequence, which are shaped primarily by the inclusion of
gravitational settling and diffusion of heavy elements. The models
diverge significantly, however, on the pre-MS, where the influence of
hydrodynamic mixing in radiative regions dominates the evolutionary
pathway while the transient convective core is at its maximum
extent. (Our theoretical treatment naturally predicts a smaller effect
on the HR diagram for stars with smaller convective cores. This is
consistent with the results of \citet{mic04}, which find minimal
overshooting for the stars near the convective/radiative boundary in
M67, a result which contradicts simple parameterized overshooting.) In
at least some cases, a good fit to observations can be achieved
without including physics which may be very important to the overall
evolution. This may be adequate for describing the state of an individual
star, but presents a serious problem for characterizing the behavior
of a star or population over time.

The eclipsing binaries provide a test of our physics, particularly the
more complete mixing, over a wide range of stellar masses. The
systematic problems with massive star models, which were identified in
\citet{ymal}, are ameliorated by the new treatment. The models are no
longer underluminous, and the central condensations as measured by
apsidal motions are no longer too small. Both of these improvements
arise from larger convective core sizes resulting from the improved
mixing. Simultaneously, the fits for almost all of the lower mass
stars improve as well. The one case in which the error formally increases
varies within the observational errors. All of the poor or
marginal fits in \citet{ymal} improve dramatically. Some of the
improvement in these lower mass models arises from the use of
non-solar abundances in a few cases, but the composition alone cannot
account for all of the error in the earlier models. AI Hya is a
particularly fine example, for the higher mass star lies redward of
the TAMS in models with incomplete mixing physics, a situation which
metallicity cannot help. More mixing is
necessary for good agreement. We do not optimize our models by varying
composition. It is changed only when an abundance estimate appears in
the literature.

Further insight into potential pitfalls can be found in the binary
sample. An increase in metallicity moves the tracks in the opposite
sense of more complete mixing. It is possible to achieve an equally
good fit with low metallicity and incomplete mixing or higher
metallicity and more realistic physics. The practice of making stellar
abundance determinations by fitting evolutionary tracks is dangerous
unless the physics in the code is very well tested independently. {\em
It is vital that accurate spectroscopic abundance determinations be
made for stars used as test cases of stellar evolution, particularly
eclipsing binaries.} 

The pre-MS systems identify an area where our physics is still
inadequate. Our predictive accuracy for these systems is not
satisfactory. We must make an {\it ad hoc} adjustment to the mixing
length in order to get large enough model radii. This tells us that
our description of convection is insufficiently physical. Further
multidimensional simulations of envelope convection in low surface
gravity stars is necessary to resolve this problem.

When coupled with the observational tests of light element depletion
and turnoff ages in young clusters in \citet{ykra} we explore the
performance of TYCHO on stars with both convective and radiative cores
and convective envelopes of various sizes on the pre-MS and main
sequence. All of these tests are performed with the same physics. No
changes are made to the mixing or composition in order to improve our
agreement with the observations. We find a strong increase in the
predictive (as opposed to calibrated) accuracy throughout this range
of conditions. This is of course a small sub-set of problems in
stellar astrophysics, but we have increasing confidence in extending
the approach to problems in other areas of stellar evolution. In the
future we plan to examine the impact of this new approach on AGB
nucleosynthesis, nucleosynthesis in very low metallicity and evolved
massive stars, and the evolution of extremely massive stars which
become luminous blue variables and SNIb/c progenitors.

\begin{acknowledgements}
This work was supported in part by the DOE (DE-FG03-98DP00214/A001 and
the ASCII FLASH center at the University of Chicago). We wish to thank
J. Liebert, E. Mamajek, C. Meakin, and C. Charbonnel for helpful
discussions. We would especially like to thank our scientific editor
for his efforts in getting this paper refereed and published in a
timely fashion.
\end{acknowledgements}

\clearpage

\begin{deluxetable}{crrllllll}
\tabletypesize{\scriptsize}
\tablecaption{Observed parameters for selected binary 
systems.\tablenotemark{a} \label{tbl-3}}
\tablewidth{0pt}
\tablehead{
\colhead{System} & \colhead{P(d)}   & \colhead{Star}   &
\colhead{Spect.} & \colhead{Mass/\sol} & \colhead{Radius/$R_\odot$} &
\colhead{$\log g{\rm (cm/s^2)}$}     & \colhead{$\log T_e{\rm(K)}$}  &
\colhead{$\log L/L_\odot$}           
}
\startdata
EM Car & $3.41$ & A & O8V  & $22.3\pm 0.3$\tablenotemark{b} &$9.34\pm 0.17$  &
 $3.864\pm 0.017 $\tablenotemark{b} & $4.531\pm 0.026$ & $5.02\pm 0.10 $  \\
HD97484 & \nodata & B & O8V  & $20.3\pm 0.3$\tablenotemark{b} &$8.33\pm 0.14$ &
 $3.905\pm 0.016 $\tablenotemark{b} & $4.531\pm 0.026$ & $4.92\pm 0.10 $   \\

V478 Cyg & $2.88$ & A & O9.5V  & $16.67\pm 0.45$ &$7.423\pm 0.079$  &
 $3.919\pm 0.015 $ & $4.484\pm 0.015$ & $4.63\pm 0.06 $  \\
HD193611 & \nodata & B & O9.5V  & $16.31\pm 0.35$ &$7.423\pm 0.079$  &
 $3.909\pm 0.013 $ & $4.485\pm 0.015$ & $4.63\pm 0.06 $   \\

CW Cep & $2.73$ & A & B0.5V &  $12.9\pm 0.1$\tablenotemark{c} & $5.685\pm 0.130$  &
 $4.039\pm 0.024$\tablenotemark{c} & $4.449\pm 0.011$\tablenotemark{d} & $4.26\pm 0.06$\tablenotemark{e} \\
HD218066 & \nodata & B & B0.5V &$11.9\pm 0.1$\tablenotemark{c}  & $5.177\pm 0.129$ &
$ 4.086\pm 0.024 $\tablenotemark{c} & $4.439\pm 0.011 $\tablenotemark{d} & $4.14\pm 0.07 $\tablenotemark{e} \\
\\
QX Car & $4.48$ & A & B2V &  $9.267\pm 0.122$ & $4.289\pm 0.091$  &
 $4.140\pm 0.020$ & $4.395\pm 0.009$\tablenotemark{d} & $3.80\pm 0.04$\tablenotemark{e} \\
HD86118 & \nodata & B & B2V &$8.480\pm 0.122$  & $4.051\pm 0.091$ &
$ 4.151\pm 0.021 $ & $4.376\pm 0.010 $\tablenotemark{d} & $3.67\pm 0.04 $\tablenotemark{e} \\

CV Vel & $6.89$ & A & B2.5V &  $6.100\pm 0.044$ & $4.087\pm 0.036$  &
 $4.000\pm 0.008$ & $4.254\pm 0.012$\tablenotemark{d} & $3.19\pm 0.05$ \\
HD77464 & \nodata & B & B2.5V &$5.996\pm 0.035$  & $3.948\pm 0.036$ &
$ 4.023\pm 0.008 $ & $4.251\pm 0.012 $\tablenotemark{d} & $3.15\pm 0.05 $ \\

U Oph & $1.68$ & A & B5V &  $5.198\pm 0.113$ & $3.438\pm 0.044$  &
 $4.081\pm 0.015$ & $4.211\pm 0.015$\tablenotemark{d} & $2.87\pm 0.08$\tablenotemark{e}\\
HD156247 & \nodata & B & B6V &$4.683\pm 0.090$  & $3.005\pm 0.055$ &
$ 4.153\pm 0.018 $ & $4.188\pm 0.015$\tablenotemark{d} & $2.66\pm 0.08 $\tablenotemark{e} \\
\\
$\zeta$ Phe & $1.67$ & A & B6V & $3.930\pm 0.045$ & $2.851\pm 0.015$ & 
$4.122\pm 0.009$ & $4.149\pm 0.010$\tablenotemark{d} & $2.46\pm 0.04$\tablenotemark{e} \\
HD6882 & \nodata & B & B8V & $2.551\pm 0.026$ & $1.853\pm 0.023$ & 
$4.309\pm 0.012$ & $4.072\pm 0.007$\tablenotemark{d} & $1.78\pm 0.04$\tablenotemark{e} \\

IQ Per & 1.74 & A & B8V & $3.521\pm 0.067$ & $2.446\pm 0.026$ & 
$4.208\pm 0.019$ & $4.111\pm 0.008$\tablenotemark{d} & $2.17\pm 0.03$\tablenotemark{e} \\
HD24909 & \nodata & B & A6V & $1.737\pm 0.031$ & $1.503\pm 0.017$ & 
$4.323\pm 0.013$ & $3.906\pm 0.008$\tablenotemark{d} & $0.93\pm 0.04$\tablenotemark{e} \\

PV Cas & 1.75 & A & B9.5V & $2.815\pm 0.050$\tablenotemark{d} & $2.297\pm 0.035$\tablenotemark{d} & 
$4.165\pm 0.016$\tablenotemark{d} & $4.032\pm 0.010$\tablenotemark{d} & $1.80\pm 0.04$\tablenotemark{e} \\
HD240208 & \nodata & B & B9.5V & $2.756\pm 0.054$\tablenotemark{d} & $2.257\pm 0.035$\tablenotemark{d} & 
$4.171\pm 0.016$\tablenotemark{d} & $4.027\pm 0.010$\tablenotemark{d} & $1.77\pm 0.04$\tablenotemark{e} \\
\\
AI Hya & 8.29 & A & F2m & $2.145\pm 0.038$ & $3.914\pm 0.031$ & 
$3.584\pm 0.011$ & $3.851\pm 0.009$\tablenotemark{d} & $1.54\pm 0.02$\tablenotemark{e} \\
+0\degr 2259 & \nodata & B & F0V & $1.978\pm 0.036$ & $2.766\pm 0.017$ & 
$3.850\pm 0.010$ & $3.869\pm 0.009$\tablenotemark{d} & $1.31\pm 0.02$\tablenotemark{e} \\

VV Pyx & 4.60 & A & A1V & $2.101\pm 0.022$ & $2.167\pm 0.020$ & 
$4.089\pm 0.009$ & $3.979\pm 0.009$\tablenotemark{d} & $1.54\pm 0.04$ \\
HD71581 & \nodata & B & A1V & $2.099\pm 0.019$ & $2.167\pm 0.020$ & 
$4.088\pm 0.009$ & $3.979\pm 0.009$\tablenotemark{d} & $1.54\pm 0.04$ \\

RS Cha & 1.67 & A & A8V   & $1.858\pm 0.016$ & $2.137\pm 0.055$ &
$4.047\pm 0.023$ & $3.883\pm 0.010$\tablenotemark{d} & $1.14\pm 0.05$\tablenotemark{e} \\
HD75747  & \nodata & B & A8V  & $1.821\pm 0.018$ & $2.338\pm 0.055$ &
$3.961\pm 0.021$ & $3.859\pm 0.010$\tablenotemark{d} & $1.13\pm 0.05$\tablenotemark{e} \\
\\
EK Cep & 4.43 & A & A1.5V & $2.029\pm 0.023$ & $1.579\pm 0.007$ & 
$4.349\pm 0.010$ & $3.954\pm 0.010$ & $1.17\pm 0.04$ \\
HD206821 & \nodata & B & G5Vp & $1.124\pm 0.012$ & $1.320\pm 0.015$ & 
$4.25\pm 0.010$ & $3.756\pm 0.015$ & $0.19\pm 0.07$ \\

TZ For &75.67 & A & F6IV  & $1.949\pm 0.022$ & $3.96\pm 0.09$ & $3.532\pm 0.02$ & $3.803\pm 0.007$ & $1.36\pm 0.06$ \\
HD20301 & \nodata & B & G8III  & $2.05\pm 0.06$ & $8.32\pm 0.12$ & $2.91\pm 0.017$ & $3.699\pm 0.009$ & $1.59\pm 0.04$ \\

MY Cyg & 4.01 & A & F0m & $1.811\pm 0.030$ & $2.193\pm 0.050$ & 
$4.007\pm 0.021$ & $3.850\pm 0.010$\tablenotemark{d} & $1.03\pm 0.04$\tablenotemark{e} \\
HD193637 & \nodata & B & F0m & $1.786\pm 0.025$ & $2.193\pm 0.050$ & 
$4.014\pm 0.021$ & $3.846\pm 0.010$\tablenotemark{d} & $1.02\pm 0.04$\tablenotemark{e} \\
\\
PV Pup & 1.66 & A & A8V & $1.565\pm 0.011$ & $1.542\pm 0.018$ & 
$4.257\pm 0.010$ & $3.870\pm 0.01$\tablenotemark{g} & $0.81\pm 0.08$\tablenotemark{e} \\
HD62863 & \nodata & B & A8V & $1.554\pm 0.013$ & $1.499\pm 0.018$ & 
$4.278\pm 0.011$ & $3.870\pm 0.01$\tablenotemark{g} & $0.79\pm 0.08$\tablenotemark{e} \\

DM Vir\tablenotemark{f} & 4.67 & A & F7V & $1.454\pm 0.008$ & $1.763\pm 0.017$ & 
$4.108\pm 0.009$ & $3.813\pm 0.007$ & $0.67\pm 0.03$\tablenotemark{h} \\
HD123423\tablenotemark{f} & \nodata & B & F7V & $1.448\pm 0.008$ & $1.763\pm 0.017$ & 
$4.106\pm 0.009$ & $3.813\pm 0.020$ & $0.67\pm 0.03$\tablenotemark{h} \\

V1143 Cyg & 7.64 & A & F5V & $1.391\pm 0.016$ & $1.346\pm 0.023$ & 
$4.323\pm 0.016$ & $3.820\pm 0.007$\tablenotemark{d} & $0.49\pm 0.03$\tablenotemark{e} \\
HD185912 & \nodata & B & F5V & $1.347\pm 0.013$ & $1.323\pm 0.023$ & 
$4.324\pm 0.016$ & $3.816\pm 0.007$\tablenotemark{d} & $0.46\pm 0.03$\tablenotemark{e} \\
\\
UX Men & 4.18 & A & F8V & $1.238\pm 0.006$ & $1.347\pm 0.013$ & 
$4.272\pm 0.009$ & $3.789\pm 0.007$\tablenotemark{g} & $0.38\pm 0.03$\tablenotemark{e} \\
HD37513 & \nodata & B & F8V & $1.198\pm 0.007$ & $1.274\pm 0.013$ & 
$4.306\pm 0.009$ & $3.781\pm 0.007$\tablenotemark{g} & $0.32\pm 0.03$\tablenotemark{e} \\
\\

\enddata
\tablenotetext{a}{Detailed references and discussion may be 
found in \citep{and91}.}
\tablenotetext{b}{\citet{slc94}.}
\tablenotetext{c}{\citet{skp92}.}
\tablenotetext{d}{\citet{rib00}.}
\tablenotetext{e}{Adjusted here for new $T_{eff}$ and $R$ determinations.}
\tablenotetext{f}{\citet{lat96}.}
\tablenotetext{g}{\citet{las02}.}
\tablenotetext{h}{\citet{hw04}.}
\end{deluxetable}

\begin{deluxetable}{crrrrrrrr}
\tabletypesize{\scriptsize}
\tablecaption{Results for selected binary systems. \label{tbl-4}}
\tablewidth{0pt}
\tablehead{
\colhead{System} & \colhead{Star}   & \colhead{Mass} & 
 \colhead{$\log R/R_\odot$} &
 \colhead{$\log T_e$}  &\colhead{$\log L$}     & \colhead{log Age (yr)} & 
 \colhead{$z$} & \colhead{$\chi^2$}
}
\startdata
EM Car & A & $22.30$  & 0.969 & 4.526 & 4.996 & 6.075 & 0.0189 & 0.16   \\
HD97484 & B & $20.30$  & 0.926 & 4.523 & 4.898 & 6.076 &  &        \\

V478 Cyg & A & $16.71$  & 0.881 & 4.476 & 4.618 & 6.310 & 0.0189 & 0.93   \\
HD193611 & B & $16.31$  & 0.865 & 4.474 & 4.580 & 6.312 &  &        \\

CW Cep   & A & $12.90$   & 0.764 & 4.437 & 4.228 & 6.404 & 0.0189 & 0.98  \\
HD218066 & B & $11.90$  & 0.722 & 4.424 & 4.093 & 6.406 &  &       \\
\\
QX Car   & A & 9.267     & 0.640 & 4.372 & 3.721 & 6.531 & 0.0189 & 0.30  \\
HD86118  & B & 8.480     & 0.602 & 4.354 & 3.576 & 6.563 &  &        \\

CV Vel   & A & 6.100     & 0.609 & 4.255 & 3.193 & 7.295 & 0.0189 & 0.08  \\
HD77464 & B & 5.996      & 0.602 & 4.251 & 3.159 & 7.299 &  &       \\

U Oph    & A & 5.198     & 0.535 & 4.221 & 2.906 & 7.379 & 0.0189 & 0.03  \\
HD156247& B & 4.683      & 0.484 & 4.197 & 2.708 & 7.380 &  &       \\
\\
$\zeta$\ Phe& A & 3.930  & 0.455 & 4.157 & 2.490 & 7.703 & 0.013 & 4.03 \\
HD6882  & B & 2.551      & 0.261 & 4.055 & 1.693 & 7.728 &  &       \\

IQ Per   & A & 3.521     & 0.382 & 4.124 & 2.211 & 7.547 & 0.0189 & 3.07 \\
HD24909 & B & 1.737      & 0.180 & 3.913 & 0.965 & 7.547 &  &       \\

PV Cas   & A & 2.827     & 0.348 & 4.037 & 1.797 & 6.490 & 0.0189 & 1.00  \\
HD240208& B & 2.768      & 0.357 & 4.027 & 1.772 & 6.491 &  &       \\
\\
AI Hya   & A & 2.145     & 0.602 & 3.834 & 1.492 & 8.556 & 0.023 & 4.12\\
$+0\degr 2259$ & B &1.978& 0.434 & 3.877 & 1.329 & 8.558 &  &       \\

VV Pyx   & A & 2.101     & 0.339 & 3.981 & 1.555 & 8.610 & 0.007 & 0.33 \\
HD71581 & B & 2.099      & 0.339 & 3.980 & 1.553 & 8.612 &  &       \\

RS Cha   & A & 1.858     & 0.317 & 3.903 & 1.198 & 6.866 & 0.0189 & 2.78  \\
HD75747 & B & 1.821      & 0.358 & 3.880 & 1.189 & 6.867 &  &       \\
\\
EK Cep   & A & 2.029     & 0.209 & 3.968 & 1.242 & 7.357 & 0.0189 & 5.13 \\
HD206821& B & 1.124      & 0.108 & 3.749 & 0.165 & 7.357 &  &       \\

TZ For   & A & 1.95      & 0.552 & 3.818 & 1.428 & 8.713 & 0.024  & 0.77 \\
HD20301  & B & 2.05      & 0.920 & 3.700 & 1.331 & 8.724 &  &      \\

MY Cyg   & A & 1.811     & 0.337 & 3.867 & 1.095 & 8.649 & 0.023 & 3.79  \\
HD193637& B & 1.786      & 0.327 & 3.865 & 1.066 & 9.651 &  &       \\
\\
PV Pup   & A & 1.565     & 0.184 & 3.872 & 0.810 & 8.520 & 0.0189 & 0.23  \\
HD62863 & B & 1.554      & 0.180 & 3.871 & 0.795 & 8.527 & &       \\

DM Vir   & A & 1.460     & 0.249 & 3.816 & 0.714 & 9.149 & 0.023 & 0.41  \\
HD123423& B & 1.454     & 0.243 & 3.817 & 0.706 & 9.149 &   &      \\

V1143 Cyg& A & 1.391     & 0.128 & 3.826 & 0.515 & 8.739 & 0.0189 & 1.48  \\
HD185912& B & 1.347      & 0.109 & 3.819 & 0.446 & 8.754 &  &       \\
\\
UX Men   & A & 1.238     & 0.134 & 3.795 & 0.400 & 9.542 & 0.021 & 1.56  \\
HD37513 & B & 1.198      & 0.096 & 3.795 & 0.323 & 9.544 &  &       \\

\enddata 
\end{deluxetable}

\begin{deluxetable}{rcrrrrrrr}
\tabletypesize{\scriptsize}
\tablecaption{Apsidal comparisons for selected binary systems. \label{tbl-5}}
\tablewidth{0pt}
\tablehead{
\colhead{System} & \colhead{Star}   & \colhead{Mass} & \colhead{$-\log k_i$} & 
  \colhead{$(k_2R^5)$ \tablenotemark{a}} &
\colhead{$\rm P/U_{CL}$ \tablenotemark{b}}    & 
\colhead{$\rm P/U_{GR}$ \tablenotemark{b}}    &
\colhead{$\rm P/U_{CL+GR}$ \tablenotemark{b}} &
\colhead{$\rm P/U_{OBS}$ \tablenotemark{b}} 
}
\startdata
EM Car  & A & $22.3$  & 1.920 & $347.0$ & $2.091$ & $0.275$ &
 2.37 & $2.2 \pm 0.3$ \\
        & B & $20.3$  & 1.935 & $267.5$           \\
\\
V478 Cyg  & A & $16.67$  & 1.935 & $160.6$ & $2.97 $ & $0.223$ &
3.19  & $3.0 \pm 0.3$ \\
        & B & $16.31$  & 1.917 & $140.0$           \\
\\
CW Cep & A & $12.90$ & $ 1.878$ & $49.50$ & $1.46$ & $0.178$ &
1.63    &  $1.640 \pm 0.014$ \\  
	& B & $11.90$ & $1.894$ & $32.11$ \\
\\
QX Car & A & $9.267$ & $ 1.898$ & $13.97$ & $0.156$ & $0.170$ &
0.326   &  $0.340 \pm 0.006$ \\  
	& B & $8.480$ & $1.936$ & $10.58$ \\
\\
U Oph & A & $5.198$ & $ 2.124$ & $2.687$ & $1.85$ & $0.0827$ &
1.93    &  $2.2 \pm 0.3$ \\  
	& B & $4.683$ & $2.111$ & $1.571$ \\
\\
$\zeta$ Phe & A & $3.930$ & $ 2.308$ & $1.025$ & $0.894$  & $0.0624$ &
0.956   &  $1.03 \pm 0.15$ \\  
	& B & $2.551$ & $2.333$ & $0.2018$ \\
\\
IQ Per & A & $3.521$ & $ 2.278$ & $0.4478$ & $0.335$ & $0.0553$ &
0.410   &  $0.40 \pm 0.03$ \\  
	& B & $1.737$ & $2.416$ & $0.0401$ \\
\\
PV Cas & A & $2.815$ & $ 2.149$ & $0.2012$ & $0.414$ & $0.0572$ &
0.499   &  $0.510 \pm 0.011$ \\  
	& B & $2.756$ & $2.221$ & $0.2375$ \\
\\

\enddata

\tablenotetext{a}{Radii $R$ in solar units.}
\tablenotetext{b}{Multiply tabular value by $ 10^{-4}$.}

\end{deluxetable}

\clearpage

\end{document}